# Spin transport in ferromagnet-InSb nanowire quantum devices

Zedong Yang[1], Brett Heischmidt[1], Sasa Gazibegovic[2], Ghada Badawy[2], Diana Car[2], Paul A. Crowell[1], Erik P.A.M. Bakkers[2], and Vlad S. Pribiag[1*]

[1]*School of Physics and Astronomy, University of Minnesota, MN 55455, USA*

[2]*Eindhoven University of Technology, 5600 MB, Eindhoven, The Netherlands*

**Abstract**

**Signatures of Majorana zero modes (MZMs), which are the building blocks for fault-tolerant topological quantum computing, have been observed in semiconductor nanowires (NW) with strong spin-orbital-interaction (SOI), such as InSb and InAs NWs with proximity-induced superconductivity. Realizing topological superconductivity and MZMs in this most widely-studied platform also requires eliminating spin degeneracy, which is realized by applying a magnetic field to induce a helical gap. However, the applied field can adversely impact the induced superconducting state in the NWs and also places geometric restrictions on the device, which can affect scaling of future MZM-based quantum registers. These challenges could be circumvented by integrating magnetic elements with the NWs. With this motivation, in this work we report the first experimental investigation of spin transport across InSb NWs, which are enabled by devices with ferromagnetic (FM) contacts. We observe signatures of spin polarization and spin-dependent transport in the quasi-one-dimensional ballistic regime. Moreover, we show that electrostatic gating tunes the observed magnetic signal and also reveals a transport regime where the device acts as a spin filter. These results open an avenue towards developing MZM devices in which spin degeneracy is lifted locally, without the need of an applied magnetic field. They also provide a path for realizing spin-based devices that leverage spin-orbital states in quantum wires.**

**Introduction**

Owing to a wide range of desirable properties, including large mobility[1,2], strong Rashba SOI[3] and large g-factors[4–6], InSb and InAs NWs have proven a versatile semiconductor platform for investigations of MZMs[7–12] and spin-orbit qubits[13–15]. These quantum wires can be tuned into a topological regime that hosts MZMs by means of applying an external magnetic field ($B_{\text{ex}}$). This leads to the formation of spin-helical states[16–18], in which momentum and spin become correlated, analogous to the situation in a topological insulator. Inducing superconductivity in this helical regime is then required to obtain MZMs. However, the external magnetic field limits the development of Majorana qubit architectures. First, $B_{\text{ex}}$ suppresses the superconductivity. Second, the requirement of $B_{\text{ex}} \perp B_{\text{SO}}$ (spin-orbital field) places strict constraints on any future Majorana quantum bit registers since $B_{\text{ex}}$ is applied globally. An alternative is to replace the global magnetic field by local magnetic interactions[19,20]. In this picture, the topological phase could be driven in one of several manners, including local magnetic proximity, non-equilibrium spin injection or simply local magnetic fields, obtained by integrating a ferromagnetic material with the nanowire. Specifically, spins can be injected into NWs electrically through FM contacts to break the spin degeneracy in the NW. Moreover, FMs also offer the possibility to explore the spin texture of the helical states, which cannot be accessed by spin-unpolarized probes[16–18,21], and to utilize such spin-orbital states for future quantum spin-based devices including spin filters[22,23] and modulators[24]. While past work has focused on the close integration of superconductors with InSb and InAs NWs[25–27], integration of FMs has not been reported to date. In this work, we provide a first demonstration that FMs can be successfully integrated with InSb NWs. By incorporating iron contacts to form a spin valve geometry and measuring hysteretic magneto-conductance (MC), we have observed for the first time electrical spin injection, transport and detection in a ballistic one-dimensional semiconductor-FM system. We find that for short contact spacing the NW can also act as an intrinsic spin filter.

## Experiments

To investigate magnetic effects in InSb nanowires we studied six devices (A-F) with a spin-valve geometry (two FM contacts separated by the InSb nanowire) with inter-contact spacings ranging between 200 nm and 1.2 µm, as shown in Figs. 1a,b (see the Methods section for fabrication details). The NWs used in our experiment are grown stacking-fault free in the [111] crystalline direction[1], which eliminates the Dresselhaus term, but allows for strong Rashba SOI[28]. Conductance plateaus approximately quantized at multiples of $e^2/h$ are observed in the devices with shorter contact spacing as $V_g$ is varied for $V_{sd}$ ~ 5 mV (Figs. 1d-f). These plateaus are signatures of 1D ballistic transport[6,29] in a quantum point contact (QPC). The QPC arises because the InSb NW acts as a quasi-1D waveguide in which electrons travel ballistically via a few discrete modes as a result of the lateral quantum confinement. At lower values of $V_{sd}$, the conductance exhibits oscillations superimposed on the plateaus. The periodic nature of the oscillations results in a diamond-like pattern clearly visible in the map of conductance vs. $V_{sd}$ and $V_g$ (Fig. 1c). This pattern indicates that in the lower bias regime the device acts as a Fabry-Pérot interferometer[30,31], a partially transmitting cavity for electron waves. The likely boundaries of the interferometer are energy barriers at the FM-NW interface, which cause electrons to become partially reflected multiple times and interfere with each other (see Supplementary Section 5). The transmission of the device can be tuned on or off resonance by varying $V_g$, which affects the Fermi wave-vector and the effective channel length at the Fermi level. The observation of Fabry-Pérot oscillations further demonstrates that transport is ballistic for short contact spacing and reveals that it is also phase-coherent. Note that although diamond-like conductance patterns in $V_{sd} - V_g$ also appear in quantum dots, the conductance of our devices is typically above $2e^2/h$ across the range of $V_g$ we explored, indicating that electrons are not strongly confined and that charging energies are not important.

In order to study the spin-dependent transport properties of our devices, we carried out magneto-conductance measurements. Figs. 2a,b present the conductance dependence on $V_{sd}$ and $B$, while Figs. 2 c,d illustrate the corresponding magneto-conductance change relative to the conductance at high field ($B_h$ = -100 mT), $\frac{G(B)-G(B_h)}{G(B_h)} \times 100\%$. Clear spin-valve-like hysteretic conductance peaks can be seen as differences in this ratio between the up-sweep and down-sweep panels near $B = 0$ (Figs. 2e-g), which reveal spin injection and transport of a spin-polarized current across the NW. The conductance is smallest at low source-drain bias, likely due to a small energy barrier of a ~1 meV or less at the FM-NW interface. Interestingly, the conductance ratio is much larger in this low-bias regime than at higher values of $V_{sd}$ (reaching more than 30% in magnitude), which could be due to enhanced transmission of spin-polarization from the FM into the NW owing to the interfacial barrier (see also Fig. 2h). This would be consistent with theoretical expectations for ballistic FM-semiconductor systems as well as with experiments in the diffusive limit, where it is known that rapid interfacial spin-depolarization due to the impedance mismatch between an FM metal and a semiconducting channel can be alleviated by introducing a spin-dependent energy barrier[32–34]. We note that spin transport in ballistic 1D semiconductor-FM systems has received little experimental attention to date, likely due to the challenge of obtaining suitably clean 1D semiconductor materials.

In order to better illustrate the spin-transport properties, it is useful to look at the magneto-conductance ratio, $\text{MCR} = \frac{G_\text{P} - G_\text{AP}}{G_\text{P}}$, where $G_\text{P}$ and $G_\text{AP}$ represent the conductance at the P and AP states. An inverse correlation between the MCR and both $V_\text{sd}$ and $V_\text{g}$ is observed in multiple devices. Fig. 2h compares the $V_\text{sd}$ dependence of the Device D MCR at $V_\text{g}=$ -0.75 V and 6 V, which correspond to near pinch-off and higher transmission respectively. We find that at low gate voltage, $V_\text{sd}$ has a stronger effect on tuning the MCR. To analyze the gate-dependence of the MCR in more detail, we also measured the conductance and MCR measurement as a function of $V_\text{g}$, at fixed $V_\text{sd} = 2$ mV (Fig. 2i). We find that over a large range of $V_\text{g}$, the MCR magnitude decreases, concomitant with an increase in conductance. This is consistent with lowering of the interfacial barrier by increasing either $V_\text{sd}$ or $V_\text{g}$, both cases leading to a reduced spin signal, as discussed in the previous paragraph. Other gate voltage effects may also influence spin signals. For example, the strength of Rashba SOI depends on back-gate voltage[3], which can further enhance the spin relaxation at larger $V_\text{g}$.

In addition to coarsely tuning the NW conductance, changing $V_\text{g}$ also induces finer modulations, associated with tuning of the Fabry-Pérot interference. These oscillatory features of the conductance are correlated with similar features in the MCR, such that conductance resonance peaks always correspond to a local decrease of the MCR magnitude (indicated by the dashed lines in Fig. 2i). This observation is qualitatively consistent with previous studies in carbon nanotubes[35,36] and could originate from asymmetric coupling of the NW to the two FM leads. By using a simple Fabry-Pérot interferometer model that accounts for spin, we find that the conductance oscillations can vary from being correlated to being anti-correlated with the MCR oscillations depending on the amount of asymmetry between the two lead-NW couplings (see Supplementary Section 7), sensitively reflecting changes in the quantum interference within the device.

Our results illustrate that spin imbalance can be achieved in InSb nanowires by electrical spin injection, and that the effects of the induced spin polarization can be controlled by tuning the bias and back-gate voltage.

**Origin of the hysteretic signals**

In a two-terminal geometry, charge current influenced by the switching of FMs can in principle generate hysteretic features due to the stray magnetic field[39] from the contacts or the magneto-Coulomb effect (MCE)[40–42]. To assess whether the observed features are due to spin injection and coherent transport, these other effects must be ruled out. As explained below, we find that neither the stray field effect nor the MCE cannot explain the hysteretic signals.

The stray field effect can be studied by comparing the spin-valve features with the magnetic field background[35]. Fig. 3a shows the magneto-conductance measurement on device F at different $V_g$ values. If we assume that the hysteresis is due to a switching of the stray field associated with the reversal of the leads magnetization, then this change, $\Delta B_{st}$, could be estimated from $\Delta B_{st} = \frac{\Delta G}{S}$, where $\Delta G$ is the observed hysteretic conductance change ($\Delta G$) and S is the slope (in µS/mT) of the background on a range up to ± 100 mT (dashed lines in Fig. 3a). We find that the putative $\Delta B_{st}$ exhibits large fluctuations as a function of $V_g$ (Fig. 3b), which is incompatible with the stray-field hypothesis since the fields associated with the reversal of the magnetization cannot depend on $V_g$. To obtain more information, we also compute the stray fields using micromagnetic simulations[43] (see Supplementary Section 1). The simulations show that the change in field across the semiconductor channel during the switching is less than ~100 mT, substantially less than the values of $\Delta B_{st}$, which further invalidates the stray-field hypothesis.

Further, we analyze the possibility of a MCE contribution to the observed hysteresis. MCE refers to indirect gating of the channel when the applied or stray magnetic field induce changes in the spin-split density of states in FM contacts that are capacitively coupled to the semiconductor channel[41,42]. The maximum of the MCR induced by the MCE can be estimated by equation 1. This ratio is dependent on the conductance, $G$; trans-conductance, $\frac{dG}{dV_g}$; spin polarizations, $P_{L(R)}$; coercive fields $H_{cL(R)}$; and capacitances, $C_{L,(R)}$, of the FM contacts; as well the global back-gate capacitance, $C_g$. By analyzing the evolution of the magneto-conductance as a function of $V_g$ on device D and device E, we estimate the maximum signal due to MCE as <0.1%, which is one to two orders of magnitude smaller than our MCR which is typically 2%-10% in magnitude (see Fig. 2i and Fig. S6b). The calculation details are given in Supplementary Section 6.

$$\text{MCR} = -\frac{1}{G}\frac{dG}{dV_g}\frac{g\mu_B(P_L C_L H_{cL} + P_R C_R H_{cR})}{eC_g} \quad (1)$$

The possibility of local charge effects was further investigated via the dependence on contact spacing. To achieve a meaningful comparison of the MCR over different contact spacings while mitigating the effects of device-to-device mesoscopic variations, we tuned all the devices to a very open regime, where the conductance saturates, and measured the MCR response to a small bias window (~2 mV). The inset of Fig. 3c demonstrates the decaying tendency of the hysteretic signal for contact spacings between 200 nm and 1200 nm. The MCR for the case of ballistic transport is expected to follow equation 2, where $\gamma$ is the spin asymmetry, $\tau_{sf}$ is the spin relaxation time, and $\tau_n = 2L/(v_F T)$ denotes the electron dwell time in the channel, with $L$ being the channel length, $v_F$ the Fermi velocity and $T$ the transmission coefficient at the interface[34].

$$\text{MCR} = \frac{1-\gamma^2}{\gamma^2}\frac{1}{1+\tau_n/\tau_{sf}} \quad (2)$$

To investigate the spacing-dependence of the signal, we rewrite the MCR in terms of length scales, with $l_{sf} = v_F \tau_{sf}$ being the spin-relaxation length; this predicts that the reciprocal of the MCR has a linear dependence on $L$:

$$1/\text{MCR} = \frac{\gamma^2}{1-\gamma^2} + \frac{\gamma^2}{1-\gamma^2} * \frac{2L}{Tl_{sf}} \quad (3)$$

Fig. 3c shows that a very good linear fit can be obtained for contact spacings in the range of 200nm to 1000nm and corroborates that over this range spin transport occurs in the ballistic limit, consistent with the theoretical prediction. From the fitting, we extracted a spin asymmetry $\gamma = 0.26$ and $l_{sf}$ is estimated to be 1.2 μm to 1.5 μm with $T$ in the range of 0.8 to 1 in a transparent interface limit. The extracted spin asymmetry is less than the bulk value of Fe (0.43)[44], which yields a spin-injection efficiency $\gamma/\gamma_{Fe}$ ~60%. The deviation of device F (contact spacing 1200 nm) from the linear trend is probably due to device-dependent factors which may influence the spin signal or a breakdown of the ballistic spin-transport picture as the channel length is several times larger than the mean-free path, estimated to be ~400 nm in this device by analyzing the trans-conductance[1,2].

The contact-spacing dependence, along with our discussion on stray field effects and MCE above, support the picture that the hysteretic features originate predominantly from injection and transport of spin-polarized electrons in the NW, rather than any stray-field or local charge effects. It is also noteworthy that unlike in typical spin valves, our devices have mostly negative MCR values ($G_{AP} > G_P$) and we rarely observed positive MCR, which could be a consequence of transmission effects in a quantum regime (see Supplementary Section 4 for positive MCR example and analysis on the inverse MCR signal).

**Spin-filtering effect**

Interestingly, in addition to the normal spin-valve features described above, we can also observe hysteretic signals with a qualitatively different shape by tuning the back-gate voltage. Fig. 4i shows the back-gate dependence of $G$ in device A (200 nm contact spacing). The conductance dip appearing at low $V_{sd}$ suggests the existence an effective gap in the NW subband spectrum. The value of $G$ drops from ~$2e^2/h$ to ~$e^2/h$ at the dip for the lowest bias, suggesting that approximately one subband is involved. Strikingly, when tuning the back-gate voltage into this conductance minimum ($V_g = -1.32$ V) and varying the $B$ field, the hysteresis loop becomes rectangular: rather than a conductance difference between P and AP states (as observed outside the dip, see inset of Fig. 4a), the main difference is now between the two P states corresponding to positive and negative $B_\parallel$. This suggests that the channel between the FM contacts has entered a spin filtering regime, with higher conductivity for one spin species than for the other. Figs. 4a,b show the conductance vs. $V_{sd}$ and $B$. To enhance the visibility of the spin-filtering region in the 2D plot, we normalized the conductance at each value of $V_{sd}$ to a scale of -1 to 1 (Normalized ratio = $\frac{G(B)-(G_{max}+G_{min})/2}{(G_{max}-G_{min})/2}$), where $G(B)$ is the conductance as a function of field at a certain $V_{sd}$, and $G_{max}$ and $G_{min}$ are the maximum and minimum of conductance at this $V_{sd}$ value. Figs.4c-h show the magneto-conductance at several values of $V_{sd}$. We observe a striking bias asymmetry in the spin filtering for in-plane applied magnetic fields, with the effect present for positive $V_{sd}$, while a more complicated magneto-conductance is observed for negative bias.

In the few-subband regime a quantum wire in an axial magnetic field is expected to develop a spontaneous spin polarization, which is enhanced near subband edges[45]. In our device, an axial component mainly originates from the misalignment between the total effective field and the in-plane normal direction to the nanowire (~20°) and is due to the sum of the applied field and the stray field from the contacts. The magnitude of this perpendicular component is estimated as $B_{\text{perp}} \approx 100 - 200 \text{ mT} \times \sin 20° \approx 30 - 60 \text{ mT}$. Although this model is consistent with spin-filtering, it fails to explain why the observed effect is asymmetric in the sign of $V_{\text{sd}}$ and why it is associated with a conductance dip. The earlier scenario relies on a spin-dependent transmission rate in the absence of any spin-orbit effects. However, InSb nanowires have strong spin-orbit coupling, which induces spin precession, with a characteristic spin-orbit length on the order of the contact spacing or less[3]. It is to be expected that the spin-orbit states in the nanowire experience different reflection and transmission at the InSb-Fe interface due to different hybridization between FM and semiconductor. In this scenario, spins of each species injected from the FM contacts would acquire different phases during propagation and reflection in the Fabry-Pérot interferometer, leading to spin-dependent interference and hence spin-dependent transmission probabilities, most visible in the single-subband limit. To validate this hypothesis, details such as the FM-SM interfacial hybridization, accurate evaluation of the contact barrier and inter-subband scattering are required, which are beyond the scope of this work. The combination of spin-orbit coupling and longitudinal magnetic field in a quantum wire is also expected to result in the formation of a helical gap, in which the spin and momentum are correlated [16–18,46,47]. The helical gap underpins the realization of Majorana modes in InSb and InAs nanowires[7,8], but so far has only been investigated through spin-unpolarized means[16–18]. Interestingly, the helical spin-momentum correlation should lead precisely to a spin filtering effect that depends on the sign of the source-drain bias. Similar filtering effects have been reported in studies of the helical edge states of topological insulators[48] or in the Rashba-induced-spin-splitting states (RISS) in InAs 2D electron gases[49], which can filter spins without opening a helical gap because of the spin imbalance in the bandstructure. In a semiconductor NW, generating a helical gap requires an external field perpendicular to the Rashba spin-orbit field, $\overrightarrow{B_{\text{SO}}}$, which points in the device plane and perpendicular to the nanowire axis. An estimate of a helical gap magnitude in our devices is given by $g\mu_B B_{\text{perp}} \approx$ 0.1-0.2 meV, which is much smaller than $-eV_{\text{sd}}$ (of the order of several meV), the energy scale at which the spin-filtering effect is observed. Nevertheless, even though the effective bias window is larger than the expected helical gap size, a filtering effect may still survive if the gap is within the bias window or if the RISS near the gap lead to strong spin imbalance in $k$-space[49]. However, according to this model we would expect to see the sign of $V_{\text{sd}}$ at which spin-filtering occurs to change when the direction of the effective magnetic field flips, which is absent in our observations, so the origin of the filtering effect is at present not fully elucidated.

**Conclusion and Outlook**

In conclusion, our work establishes the integration of ferromagnetic materials with InSb nanowires to achieve ballistic spin-dependent transport down to the few-subband regime. Quasi-1D ballistic transport signatures such as quantized plateaus and Fabry-Pérot interference were observed in multiple devices. The evolution of the hysteretic signals distinguishes the spin transport features from other local effects. Spin-valve effects are a generic effect observed as hysteretic magneto-conductance dependent on $V_{\text{sd}}$ and $V_{\text{g}}$ in all devices we studied down to low temperature, although the details of the hysteresis loop shapes vary from device to device, likely due to mesoscopic disorder effects. In addition, at the shortest contact spacing we observed a robust bias-asymmetric spin-filtering effect. The fact that this effect is apparent only in the shortest device, with contact spacing less than the mean free path[6,29] and spin-orbit length[3,17] reported in the literature, suggests that is associated with spin-orbit effects in the ballistic transport regime. The interplay between ferromagnetism and strong spin-orbit coupling in InSb nanowires could enable the development of Majorana zero-mode devices which can operate at zero global magnetic fields by integrating magnetic materials within the device itself. The FM-NW-FM geometry also provides an ideal platform to investigate the spin structure of the topological helical states and can enable the development of ballistic quantum devices exploiting spin-orbit effects, such as quantum spin filters and modulators. Future work will benefit from optimizing the NW-FM interface, e.g. by integration of magnetic materials grown epitaxially onto the nanowires, and by exploring the interplay between induced magnetic and superconducting properties.

## Methods:

*Device fabrication and measurement:* The devices were fabricated on p-doped Si substrate covered with 285-nm thick $SiO_2$, which serves as a global back-gate dielectric. To fully cover the NWs (~ 90nm in diameter), 120 nm thick Fe was deposited, capped with 5 nm Au (see Supplementary Section 3 for step-by-step fabrication details). The two contacts (FM1 and FM2) were designed with different geometries in order to obtain different switching fields. FM1 has dimensions 6 μm × 150 nm, with a sharp tip at one end in order to support a more uniform magnetization, while FM2 has dimensions 4.5 μm × 300 nm to obtain a slightly smaller switching field. The switching behavior of the FM contacts was simulated using the Mumax micro-magnetic simulation package and switching fields were verified by anisotropic magnetoresistance measurement of the FM leads (see Supplementary Section 1 and 2 for simulation details). All the measurements were run in an Oxford dilution refrigerator with a vector magnet and base temperature of 15 mK. In the experiments, the temperature was 500 mK (unless otherwise specified), in order to reduce the effects of mesoscopic fluctuations on the signal.

*Accounting for instrumental and contact resistances:* To obtain the actual conductance of the device, the raw data was corrected by subtracting the series filter and ammeter resistances[6,29], which combine to give $R_{\text{instrument}}$ = 4.5 kΩ + 2.1 kΩ = 6.6 kΩ. An additional device-dependent resistance, $R_{\text{dev}}$, was also subtracted to fit the first conductance plateau to the $1G_0$ level, to account for the leads' metallic resistance and the metal-semiconductor interfacial resistance[6,29]. Typical $R_{\text{dev}}$ values vary between 1 kΩ to 6 kΩ. In Figs. 1d-f, the $R_{\text{dev}}$ values used were respectively 5.5 kΩ, 1 kΩ and 3.8 kΩ. In this letter, we show the conductance in $e^2/h$ if whenever the conductance is corrected for the contacts contribution. However, as the spin signals could also be influenced by the contact resistance, we show MR values without this correction. In this case, we report the conductance in units of μS instead of $e^2/h$ to easily distinguish the two cases.

## Acknowledgements:

We acknowledge G. Graziano and Y. Ayino for help with using the dilution refrigerator, and S. Frolov, Y. Jiang, S. Heedt, X. Fu, X. Ying and P. Zhang for helpful discussions. This work was supported primarily by the Department of Energy under Award No. DE-SC-0019274. Nanowire growth was supported by the European Research Council (ERC HELENA 617256), and the Dutch Organization for Scientific Research (NWO). Portions of this work were conducted in the Minnesota Nano Center, which is supported by the National Science Foundation through the National Nano Coordinated Infrastructure Network (NNCI) under Award Number ECCS-1542202, and in the Characterization Facility, University of Minnesota, a member of the NSF-funded Materials Research Facilities Network (www.mrfn.org) via the MRSEC program.

**Author contributions**

Z.Y. and V.S.P designed the experiments. Z.Y. fabricated the devices and performed the measurements. B.H. performed the micromagnetic simulations. Z.Y., V.S.P., B.H. and P.A.C. analyzed the data. S.G., G.B., D.C. and E.P.A.M.B provided the nanowires. All authors contributed to writing the manuscript.

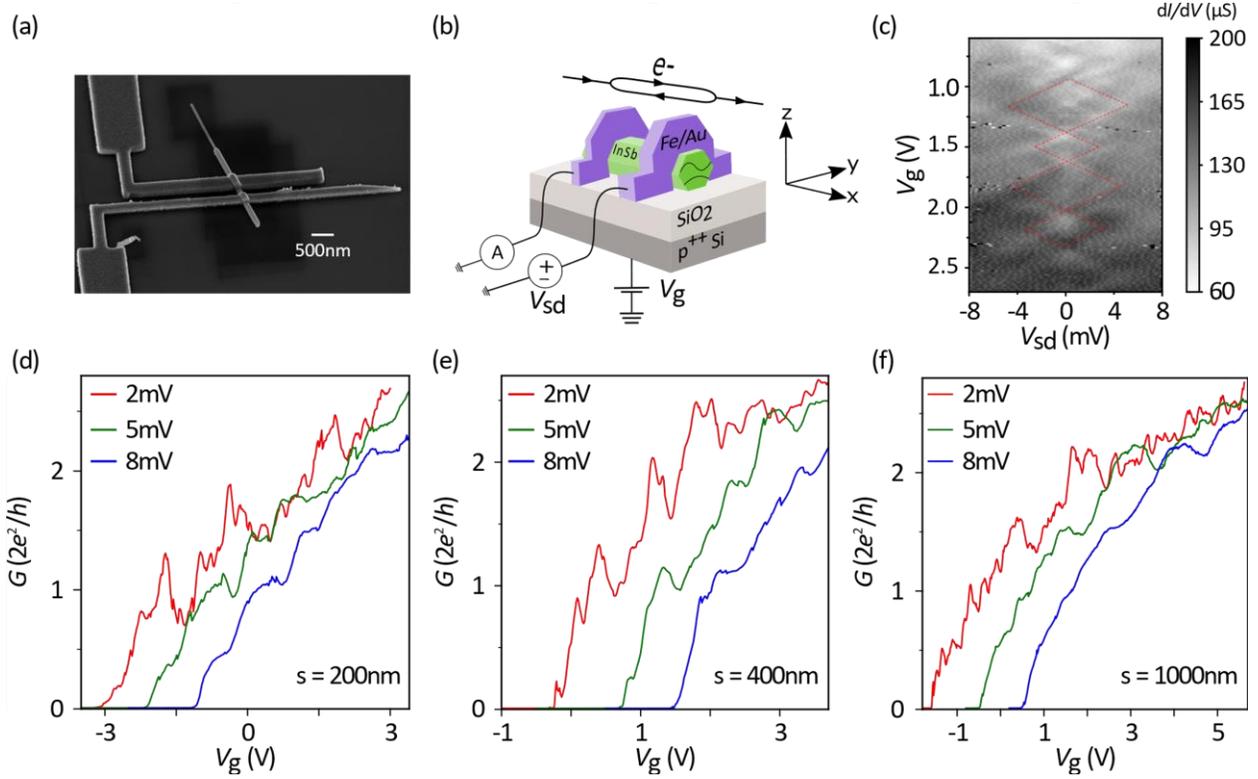

**Figure 1.** (a) False-color scanning electron micrograph of device A (spacing 200 nm). (b) Device schematic. (c) $dI/dV$ vs. $V_{sd}$ and $V_g$ (device B, spacing 400 nm) showing a diamond pattern at high $G$, a signature of Fabry-Pérot electronic interference. (d-f) DC conductance vs. back-gate voltage ($V_g$) of three separate devices A, B, E (contact spacing 200 nm, 400 nm, 1000 nm) at $V_{sd}$ of 2 mV, 5 mV and 8 mV, showing the evolution of quantized conductance plateaus with increasing $V_{sd}$ (offsets are applied on the 5 mV and 8 mV traces for clarity). Data in (d) and (e) were taken at $T = 100$ mK.

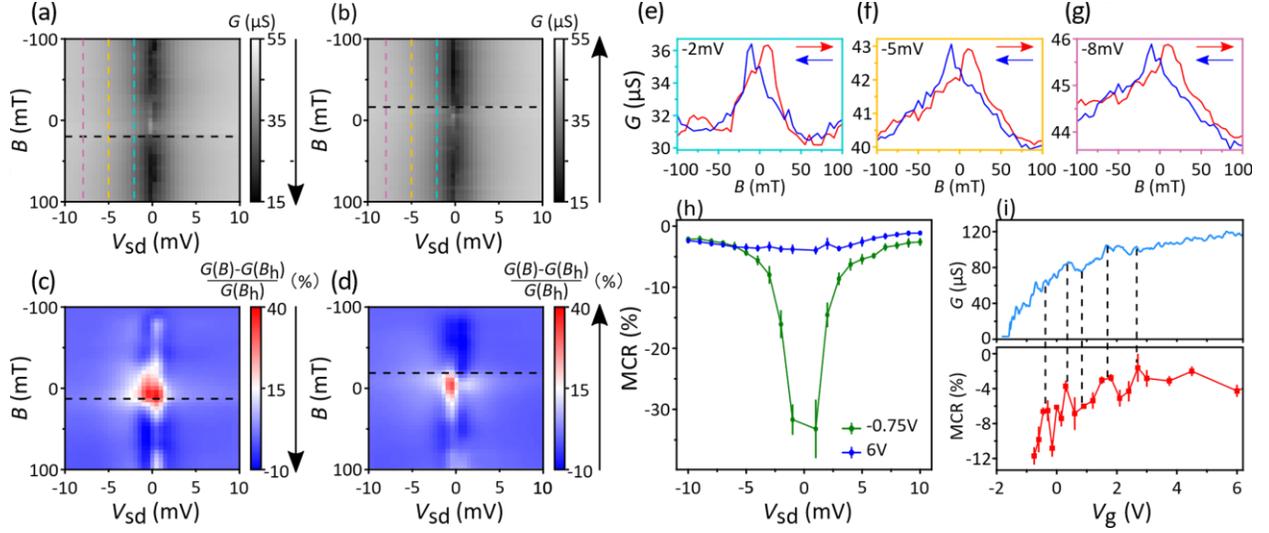

**Figure 2**. (a,b) Magneto-conductance of device E measured as a function of $V_{sd}$ and magnetic field, $B$. (c,d) Plot of $\frac{G-G(B_h)}{G(B_h)}$ to demonstrate the hysteretic features over the bias effects. The black arrows denote the field step direction and the dot lines are guide-to-eye showing where the contacts switch. (e-g) Vertical line-cuts from (a,b) at different $V_{sd}$ showing the evolution of the spin valve conductance features. Boundary colors correspond to that of dotted-lines in the 2D plots. (h) The MCR dependence on $V_{sd}$ at $V_g$ = -0.75 V and 6 V respectively. At low $V_g$, the $V_{sd}$ has stronger tunability and attains larger values. (i) Dependence of $G$ and MCR dependence on $V_g$ at fixed $V_{sd}$ (2 mV). The dash lines show the correlation between the interference-induced conductance oscillations and the MCR oscillations.

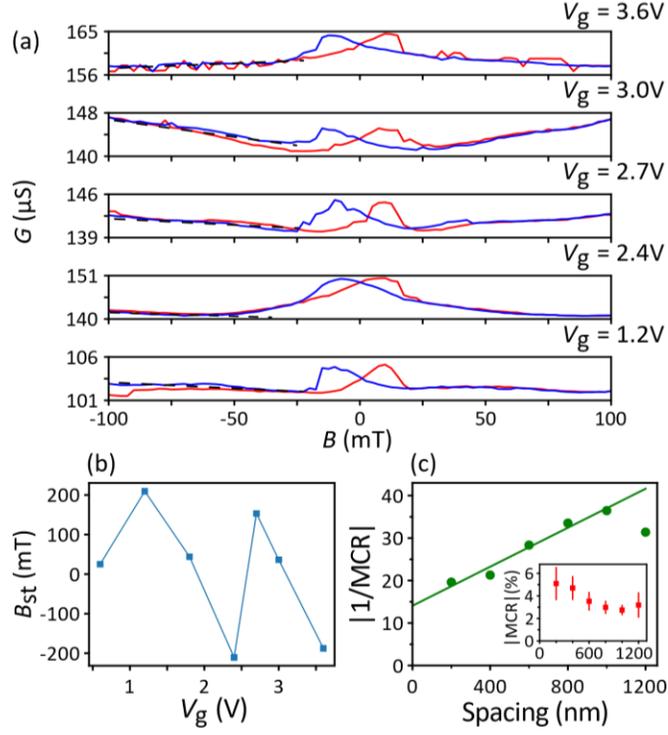

**Figure 3**. (a) Magneto-conductance of device F (spacing 1200nm) at different $V_g$. (b) Putative stray field at different $V_g$ extracted from the hysteretic signal and the background slope $\frac{\Delta G}{dG/dB}$, showing strong but random dependence on $V_g$. (c) 1/MCR as a function of contact spacing, showing linear fit to eqn. 3 (measured at a low $V_{sd} = 2$ mV). Inset: MCR vs. contact spacing.

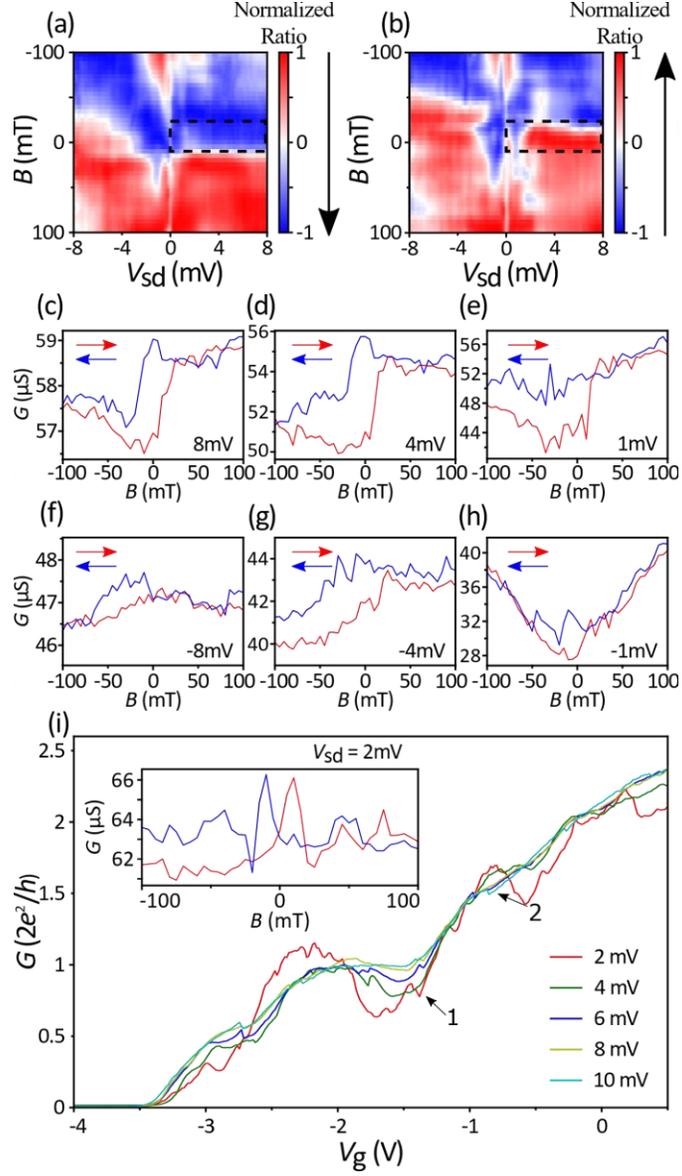

**Figure 4**. (a,b) Device A conductance measured vs. $V_{sd}$ and $B$, rescaled to a range of -1 to 1 along the magnetic field axis at $T = 100$ mK. The black arrows indicate the sweep direction. Dotted rectangle marks the region where the spin-filter effect is observed. (c-h) Line-cuts along the $B$ axis showing the spin-filter features evolution with $V_{sd}$. (i) Conductance vs. $V_g$ at different $V_{sd}$ for $B_z = 0.5$ T, $T = 100$ mK. A conductance dip is observed around $V_g = -1.5$ V, within which the spin-filtering effect is observed. Arrow 1 indicates the value of $V_g$ where the spin-filter data in (a-h) were taken. Inset: When the conductance is tuned out of the dip (arrow 2), the hysteresis loop exhibits normal spin-valve features without spin filtering.

# Supplementary Information for
# Spin transport in ferromagnet-InSb nanowire quantum devices

Zedong Yang[1], Brett Heischmidt[1], Sasa Gazibegovic[2], Ghada Badawy[2], Diana Car[2], Paul A. Crowell[1], Erik P.A.M. Bakkers[2], and Vlad S. Pribiag[1*]

[1]*School of Physics and Astronomy, University of Minnesota, MN 55455, USA*

[2]*Eindhoven University of Technology, 5600 MB, Eindhoven, The Netherlands*

## Contents:

## 1. Micromagnetic simulations

Mumax3[1] was used to simulate the Fe contacts in a realistic geometry which includes the presence of the nanowire. All proportions were taken into account: the diameter of the nanowire was 100 nm, and the thickness of the Fe was 120 nm. The experimental coercive field (~20 mT) was reproduced with the simulation, with switching fields shown in Fig.S1-1. The longer contacts were designed with the beveled tip to increase the AP magnetization states window and to obtain a cleaner switch.

The effects of an external field of 0 mT (P state), 15 mT (AP state), and 30 mT (P state in the opposite direction) were considered to explore the demagnetizing field change right before and after the magnetization switch of the two FMs (Fig.S1-2). Demagnetizing field values are shown in three regions: right underneath the two contacts and in the middle of the NW channel. The demagnetizing field in the middle height of the nanowire across the semiconductor channel has a strength of ~50 mT and right underneath the contact ~200 mT. The effective field in the nanowire is found by simply adding the applied field to the demagnetizing field. When the contacts have opposite magnetization, the direction of the effective field inside the nanowire switches direction. The values of the effective field in the NW segment between the contacts (region 2 in Fig. S1) justify considering background fields of up to ~100 mT in our analysis on the stray field effect in the main text (Fig.3b). Shown are plane cuts halfway up the wire. Plane cuts one- and two-thirds up the wire were also analyzed and the results were largely unchanged with height.

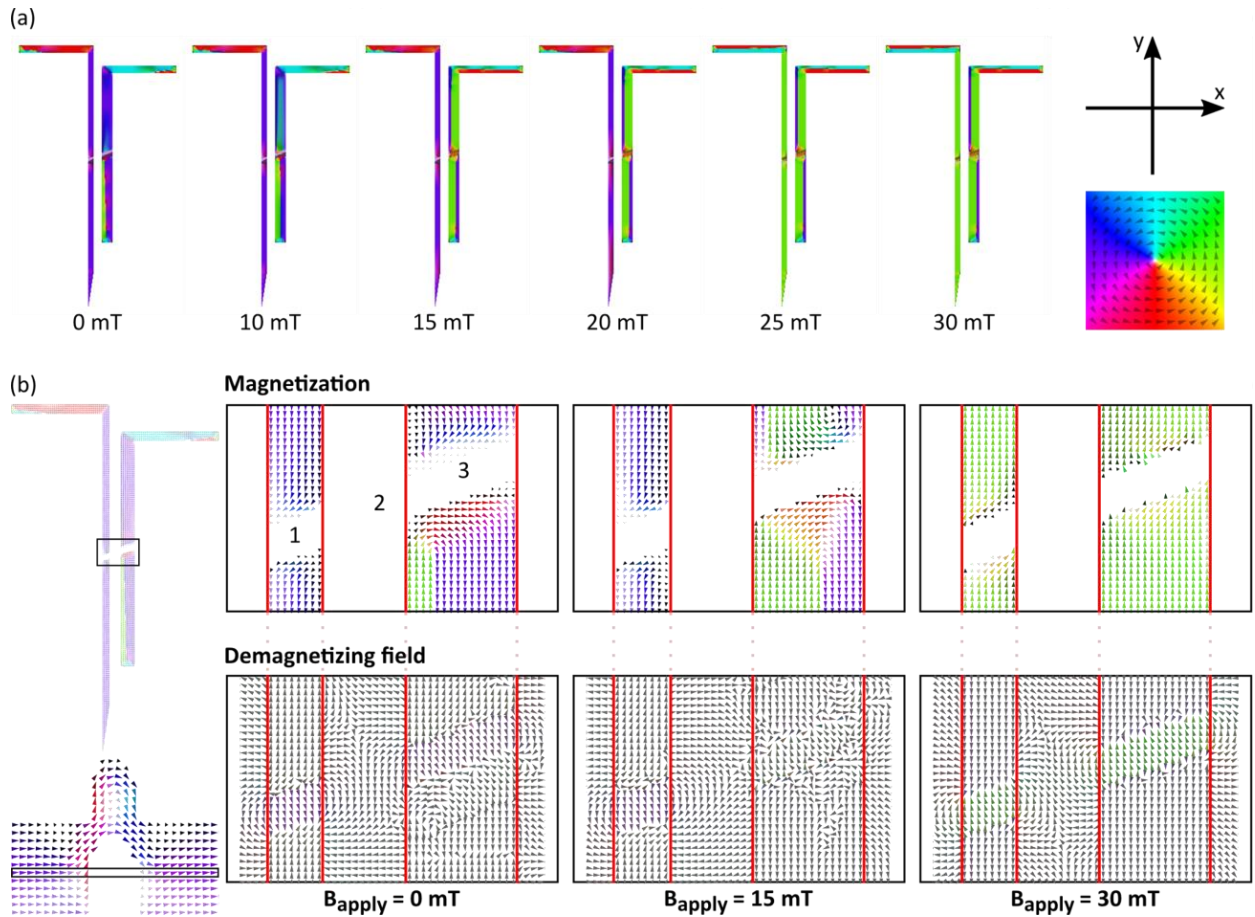

**Figure S1.** (a) Magnetic switching of contacts. (Left) Reduced magnetization of contacts with external magnetic field strengths indicated. Initial magnetization is in the -y direction and external field in +y. Switching fields are indicated in bold red type. (Right) Colors associated with direction of vector, and coordinate system. (b) Reduced magnetization and demagnetizing field in different regimes. (Left panel) Plane cuts showing zoomed area (top) and plane analyzed (bottom). (Right panels) Reduced magnetization (m) and demagnetizing field for 0 mT, 15 mT, and 30 mT applied field. The demagnetizing field values in mT for positions 1, 2, and 3 are as follows: (0 mT): -248, -44, -143; (15 mT): -186, -24, 42; (30 mT): 269, 53, 323. Red lines are guides-to-eye showing the positions of the FM contacts.

## 2. Anisotropic magnetoresistance measurement of the FM leads

In order to examine the magnetic behavior of the FM contacts we carried out anisotropic magneto-resistance[2,3] (AMR) measurements across the contacts using devices where each contact has both ends connected to a bonding pad (see device G in Fig. S2), which allows electrical measurements of the leads independent of the NW. By measuring the AMR of the FM leads, we detected the coercive field of the contacts. It is well established that the resistance of metallic ferromagnets, such as Fe, depends on the angle between the applied current direction and the magnetization orientation due to the spin-orbit coupling, which forms the basis for AMR. The AMR signal is maximum when the current and magnetization are parallel or antiparallel and minimum when at right angles to each other. At small values of the applied magnetic field, $B$, the magnetization orientation is primarily determined by the shape anisotropy and is thus largely aligned with the long axis of the FM contacts and hence predominantly parallel to the current direction everywhere within the leads. Near the coercive fields the quasi- single domain contact gradually switches its magnetization to the opposite direction as $B$ is varied, leading to regions of local magnetization that are not parallel with the contact long axis or the current. Due to AMR, the resistance is a maximum at $B = 0$ and decreases as the field changes sign due to the rotation of the magnetization. The resistance then increases abruptly at the coercive field, after which the magnetization is nearly reversed relative to its original direction. As reflected in Figs. S2 a and b, measured at $T$ ~4.2 Kelvin in a helium cryostat, the change in resistance is ~ 0.02 - 0.03 %, which is consistent with AMR measurements on iron. The hysteretic AMR dips of both the narrow and wide contacts indicate a sharp switching around 15 – 20 mT, which is consistent with the results of micromagnetic simulations described in the previous section. We also performed measurement of the same device while applying a bias across the NW (Fig. S2c). These measurements also showed hysteretic features around the same switching field observed in the AMR sweep, however the overall resistance is much higher (indicative of transport via the NW), and the magnetoresistance signal is much larger (~10%), indicating that the AMR contribution is negligible when measuring across the NW.

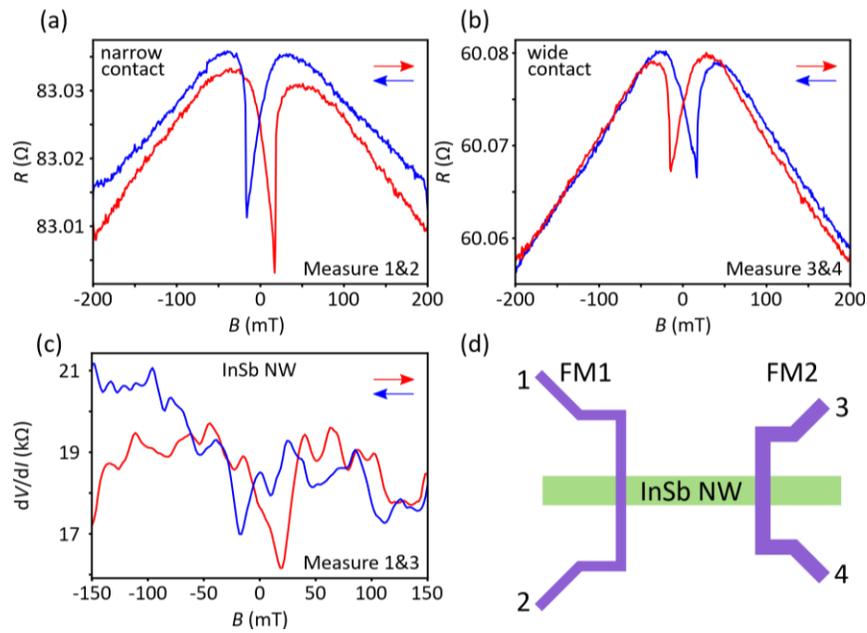

**Figure S2.** (a,b) Anisotropic magnetoresistance (AMR) measurement on the narrow lead and the wide lead, respectively (Device G). The resistance jumps demonstrate the switching of the magnetization. (c) Differential resistance measured on the same device, but through the NW shows hysteretic features at the same field values as the AMR of the FM leads. (d) Schematic of the device design and measurement.

3. **Details of fabrication process**

- Substrate: A p-doped silicon substrate covered with 285 nm-thick $SiO_2$ was used. 5 nm/50 nm Ti/Au markers were deposited to locate the NWs.

- The substrate was pre-cleaned by $O_2$ plasma and InSb NWs were subsequently transferred onto the substrate and imaged by a Keyence optical microscope.

- Resist spinning. PMMA 495K A4 was spun at 4000 RPM and baked at 180 $^o$C for ten minutes twice. Then, PMMA 950K A2 was spun at 4000 RPM and baked at 180 $^o$C for 10 minutes.

- EBeam lithography for FM leads.

- Developing in IPA : MIBK (3:1) for 80 s and rinse in IPA for 80 s at room temperature.

- RIE cleaning in AV etcher. $O_2$ plasma (power = 30W, flow = 40 sccm, time = 15s) was used to clean off resist residue.

- Sulfur passivation for contacts deposition pre-cleaning. Mixing 0.29 g sulfur powder with 3.5 mL $(NH_4)_2S$ solution then diluted with DI-$H_2O$ (1:200). Then, soaking the sample into 3 mL of the diluted solution and thermal bath at 60 $^o$C for 30 minutes.

- Contact deposition: 120 nm Fe as the spin contacts and 5 nm Au as the capping layer.

- Lift-off in acetone at 60 $^o$C for 1 hour with stirring then rinse in IPA.

## 4. The sign of the MCR

In spin valve measurements, the resistance is typically larger when the FMs are in the AP states (positive MCR), which can be explained by a two-channel resistor model[4]. In our measurement, we found that our devices have an inversed spin valve signal (negative MCR), a smaller resistance in the AP states, and we only rarely observe positive MCR. Fig. S4 demonstrates an MCR transition of device D from negative to positive. With tuning the channel slightly using the back-gate (with a conductance difference ~ 5%), the MCR transitions from negative (as observed in most of our measurements, $G_{AP} > G_P$) to positive. This transition from negative to positive suggests that interference partially determines the sign of the MCR.

Reversed spin valve signal has been reported in various experiments and its origins can be ascribed to different spin asymmetry ratio of injector and detector[5], electronic interference[6,7], interfacial band profile dependence of tunneling[8], spin-dependent hybridization between the bound states in the channel and FMs[9], and spin blockade by localized states[10].

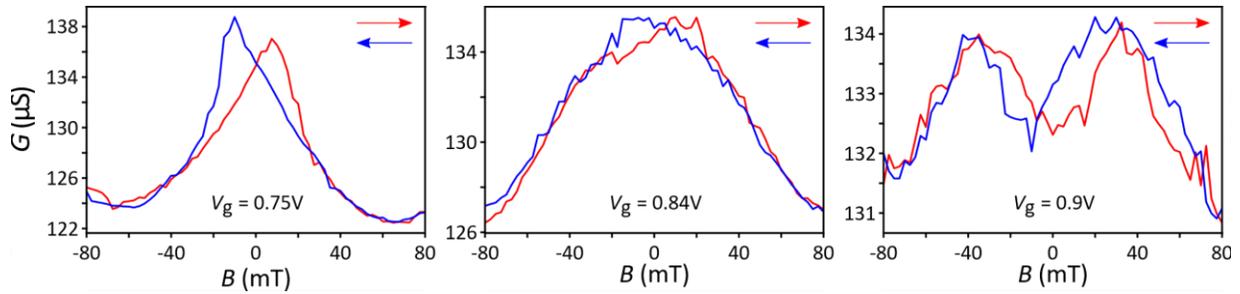

**Figure S4.** $G$ vs. $B$ measurements on device D with a fixed $V_{sd} = 2$ mV at different $V_g$ showing a sign transition of the MCR from negative to positive. (a) $V_g = 0.75$ V. (b) $V_g = 0.84$ V. (c) $V_g = 0.9$ V.

## 5. Quantum interference patterns of different devices

The Fabry-Pérot quantum interference patterns were observed as diamond shapes in the $V_{sd}$ vs $V_g$ conductance scan (Fig. 1 and Fig. S5). The presence of the interference pattern is ascribed to an energy barrier that forms under the metallic contacts. This may be in part a Schottky barrier, as expected for a metal-semiconductor interface. In addition, metallic contacts screen the electrical field, such that the back-gate tunability on the channel close to the contacts is reduced, resulting in different electron densities in that area than in-between the contacts. As a result of these barriers, electrons experience multiple partial reflections near the contacts while propagating phase-coherently, which gives rise to the Fabry-Pérot interference and quasi-bound states. By tuning the wavevector through varying $V_g$, the wave interference can be tuned from constructive to destructive, thus modulating transmission and hence the device conductance. Figs. S5 a-d display the interference patterns of devices A, D, E and F, with spacing of 200 nm, 800 nm, 1000 nm and 12 00nm respectively.

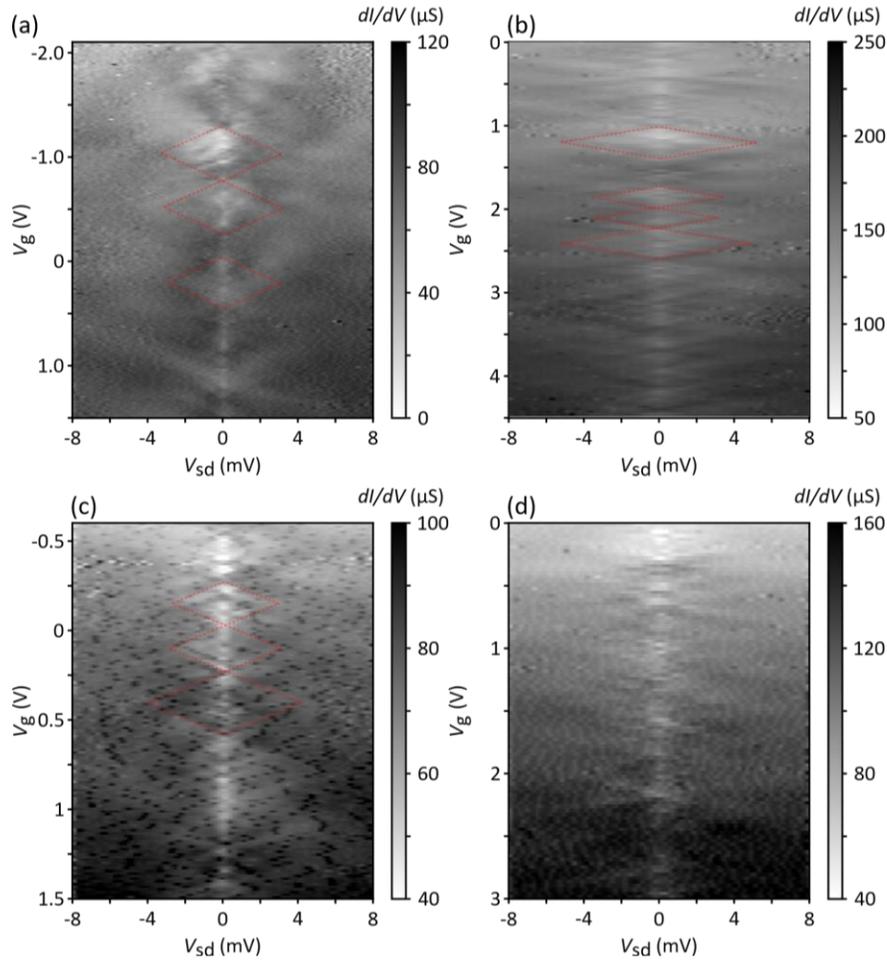

**Figure S5.** (a-d) Numerical differential conductance in vs. $V_{sd}$ and $V_g$ measured on devices A, D, E, F. Red dash lines in device A, D, E (spacing 200 nm, 800 nm and 1000 nm) are guides to the eye showing quantum interference diamond patterns. Patterns in device F are not quite clear.

## 6. Calculation of magneto-Coulomb effect-induced signal

Switching of the orientation of the magnetization in the FM contacts can also in principle induce a hysteretic magnetoconductance[11–13] via the magneto-Coulomb effect (MCE). An applied magnetic field shifts the spin energy in the contacts through the Zeeman effect. Because of the imbalanced spin density-of-states inherent in a FM, the effective Fermi energy in the FM must be changed by $\Delta\mu = -\frac{1}{2}Pg\mu_B B$ to keep the number of electrons constant. If the contacts couple capacitively to the conductive channel, this Fermi energy change translates into an equivalent shift of the back-gate potential, changing the conductance from $G(V_g)$ to $G(V_g - \Delta V)$. The MCE signal ratio can be estimated[12] by $\mathrm{MR} = -\frac{1}{G}\frac{dG}{dV_g}\frac{g\mu_B(P_L C_L H_{cL} + P_R C_R H_{cR})}{eC_g}$, where $P_{L(R)}$, $C_{L(R)}$, $H_{cL(R)}$ are the spin-polarization, capacitance and coercive field for the left and right contacts and $C_g$ is the back-gate capacitance. For simplicity (referring to the method used in ref. S16[14]), we assume symmetric parameters for the two contacts:

$$\mathrm{MR} = -\frac{1}{eG}\frac{dG}{dV_g}g\mu_B P H_c \frac{C_L + C_R}{C_g} \quad (1)$$

Here, we investigated the possible effect of MCR on Device D at small values of $V_g$ and on device E at larger values of $V_g$. Fig. S6a shows the conductance measurement of device D at 80 mT (FM contacts have parallel magnetizations) for $V_g$ between 0.6 V to 1.2 V. By extracting the maximum of the $G(V_g)$ slope, $\frac{dG}{GdV_g}$ is estimated to be ~0.63 V$^{-1}$. The spin polarization $P$ is set to be the largest possible value, 0.5, for the Fe-semiconductor interface, and the coercive field is taken to be 20 mT, consistent with our previous characterization. The capacitance ratio can be estimated by the slope $\frac{dV_g}{dV_{sd}}$ of the diamond patterns in the $V_{sd} - V_g$ scan, which is ~110 (Fig. S5b). Therefore, the maximum of the MCE induced MCR is about 0.008%, which is obviously much smaller than the corresponding MCR in figure S6b. Repeating the same process on device E (see Fig. 2i for conductance and MCR, and Fig. S5c for the $V_{sd} - V_g$ conductance scan), we obtained a maximum value of 0.08% for the MCE contribution, again much smaller than the MCR magnitude observed in Fig. 2i which typically varies from 2% to 30%. Both of the results verify that MCE is not likely to give rise to the MCR signal we observed.

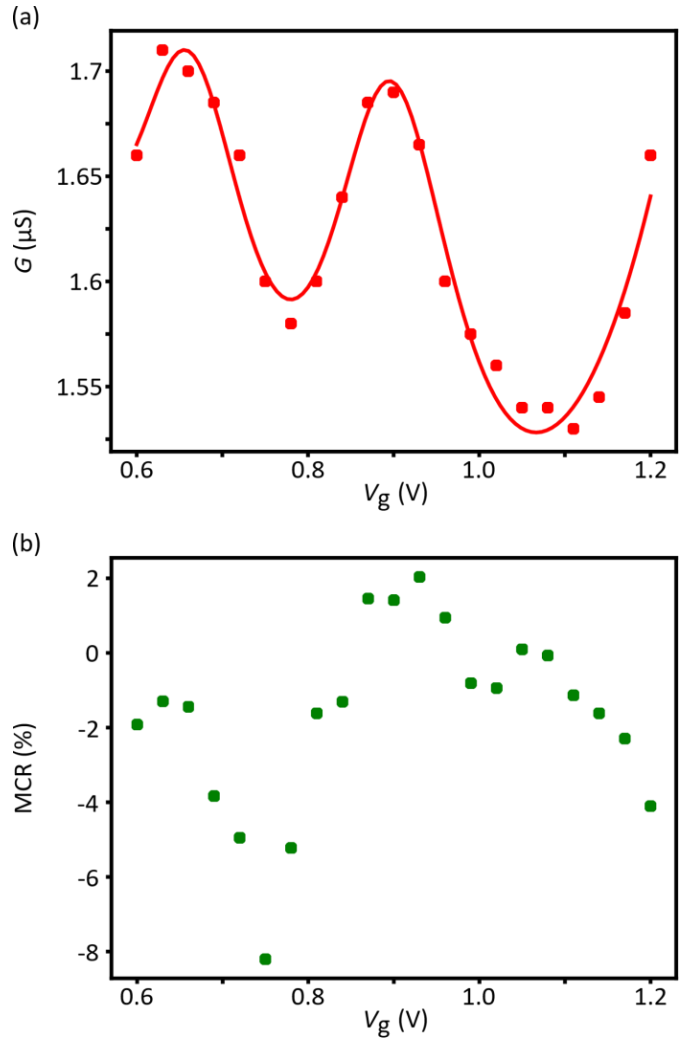

**Figure S6.** (a) Conductance measurement on device D at $B = -80$ mT with $V_{sd} = 2$mV as a function of the back-gate voltage shows an oscillating behavior, which indicates that $dG/dV_g$ can change sign. (b) The corresponding MCR at each $V_g$ shows no correlation with $dG/dV_g$.

## 7. Model for oscillating conductance

Ref. S9[7] uses a Fabry-Pérot transmission model based on the Landauer-Büttiker formalism to investigate the interference effects on the spin transport in carbon nanotubes (CNT):

$$T^{ss'} = \frac{T_L^s T_R^{s'}}{\left|1-\left[(1-T_L^s)(1-T_R^{s'})\right]^{-1/2} e^{2i\delta}\right|^2} \quad (2)$$

Here, $T_{L(R)}$ are the electronic transmission rate at the left (right) contacts, and $T_{L(R)}^{s(s')}$ is the spin-dependent transmission rate, $T_{L(R)}^s = T_{L(R)}(1+sP)$, where $P$ is the contact spin polarization and $s = \pm 1$ is the spin polarity. $\delta$ is the phase electrons acquire from propagation and reflection during one propagation cycle, which can be controlled experimentally by $V_g$. The transmission yields a channel conductance $G = \frac{e^2}{h}\sum_{ss'} T_i^{ss'}$. For simplicity, the reflection contribution was omitted. Here, we also utilize this model to study the observed correlations between oscillations in $G$ and MCR.

We looked at the contact transmission asymmetry influence on $G$ and the MCR. Barriers at the FM-InSb interfaces can be different due to the NW inhomogeneity, contacts geometry or effects of the fabrication process. Refs. S8,S9[6,7] reported that this asymmetry affects the spin signal and even reverses the sign of the magnetoresistance ratio in CNTs. Here, we calculate the conductance and MCR of a single mode by fixing one contact transmission at $T_L = 0.8$, and tuning the other $T_R$ from 0.8 to 0.2, which tunes between symmetric and asymmetric transmission. The contact polarization of an iron-semiconductor interface can be estimated based on previous Fe-GaAs study[8]. We assume the effective polarization $P$, which is equivalent to the $\gamma$ in the main text, could vary from 0.1 to 0.25. In Fig.S7 a-h we show the $G - \delta$ and MCR $- \delta$ evolution as a changing $T_R$ for $P = 0.25$ (high polarization) and $P = 0.1$ (low polarization). In both cases, we can observe a transition of the $G -$ MCR from correlation to anti-correlation as the $T_R$ is tuned to be more asymmetric. In Fig.2i and Fig.S7 d,h, our $G$ vs. $V_g$ and MCR vs. $V_g$ measurements show an anti-correlated behavior, which suggests an asymmetric transmission existing at the two contacts.

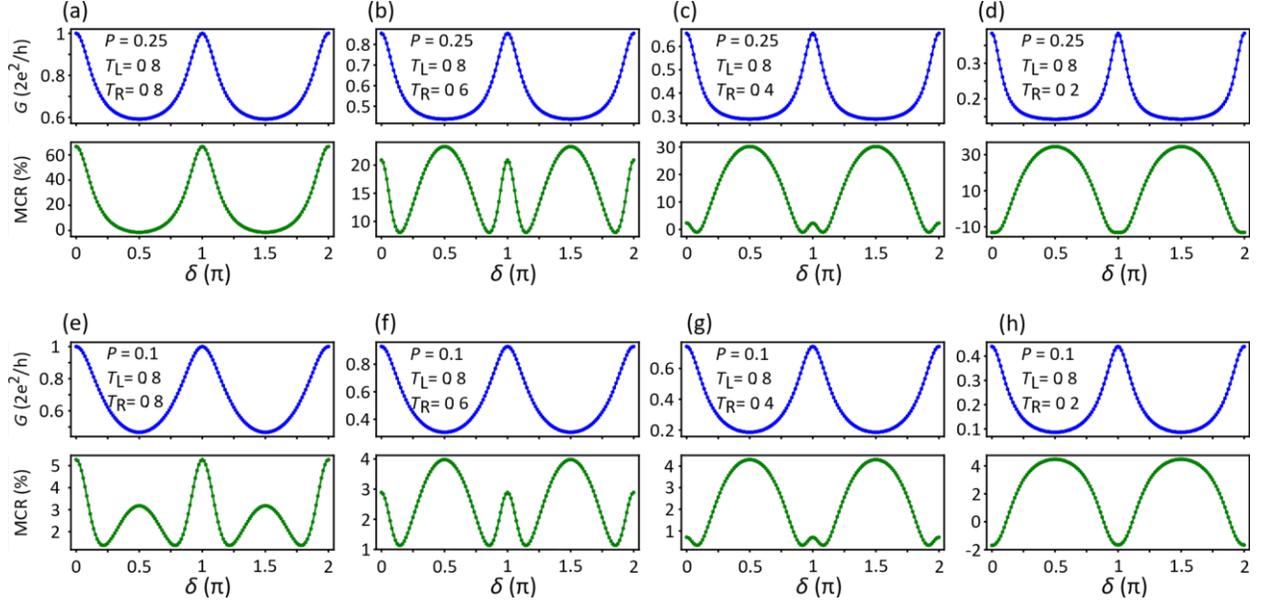

**Figure S7.** Correlation between the conductance (top row, blue curves) and the MCR (bottom row, green curves) of a single mode in a Fabry-Pérot interferometer. (a-d) $P = 0.25$, $T_L = 0.8$, $T_R$ is tuned from 0.8 to 0.2 (from symmetric barriers to asymmetric). (e-h) Same transmission parameters with $P = 0.1$.

## 8. Reproducibility of the spin-filter (SF) effect

Additional line-cuts extracted from Fig.4 a,b are displayed in Fig. S8a-q. The evolution of the filtering features clearly illustrates a bias-asymmetry, with the phenomenon being more pronounced at the positive bias side. To further explore the reproducibility of the SF signal, we also performed magnetic field sweeps (rather than the step-hold approach used to acquire the data in Fig. 4) at a fixed $V_{sd} = 2$mV. Fig. S8q illustrates the outcome of this measurement. Importantly, $V_g$ was first cycled several times between the acquisition of the data for Fig. 4 and these data, and the SF effect persisted. We also performed a thermal cycle (warm up to above 100K then re-cool down), then repeated the measurement. After the thermal cycle (Fig. S8r), we found that although the whole $G$ vs. $V_g$ trace shifted slightly along $V_g$ space, the conductance dip and associated SF effect within it persisted. This suggests that the filtering effect is quite robust. [Note: the small vertical offset between the conductance of the two sweeps at 100 mT in Fig. S8r is likely due to charge switches in the device over the measurement time scale (~40 minutes)].

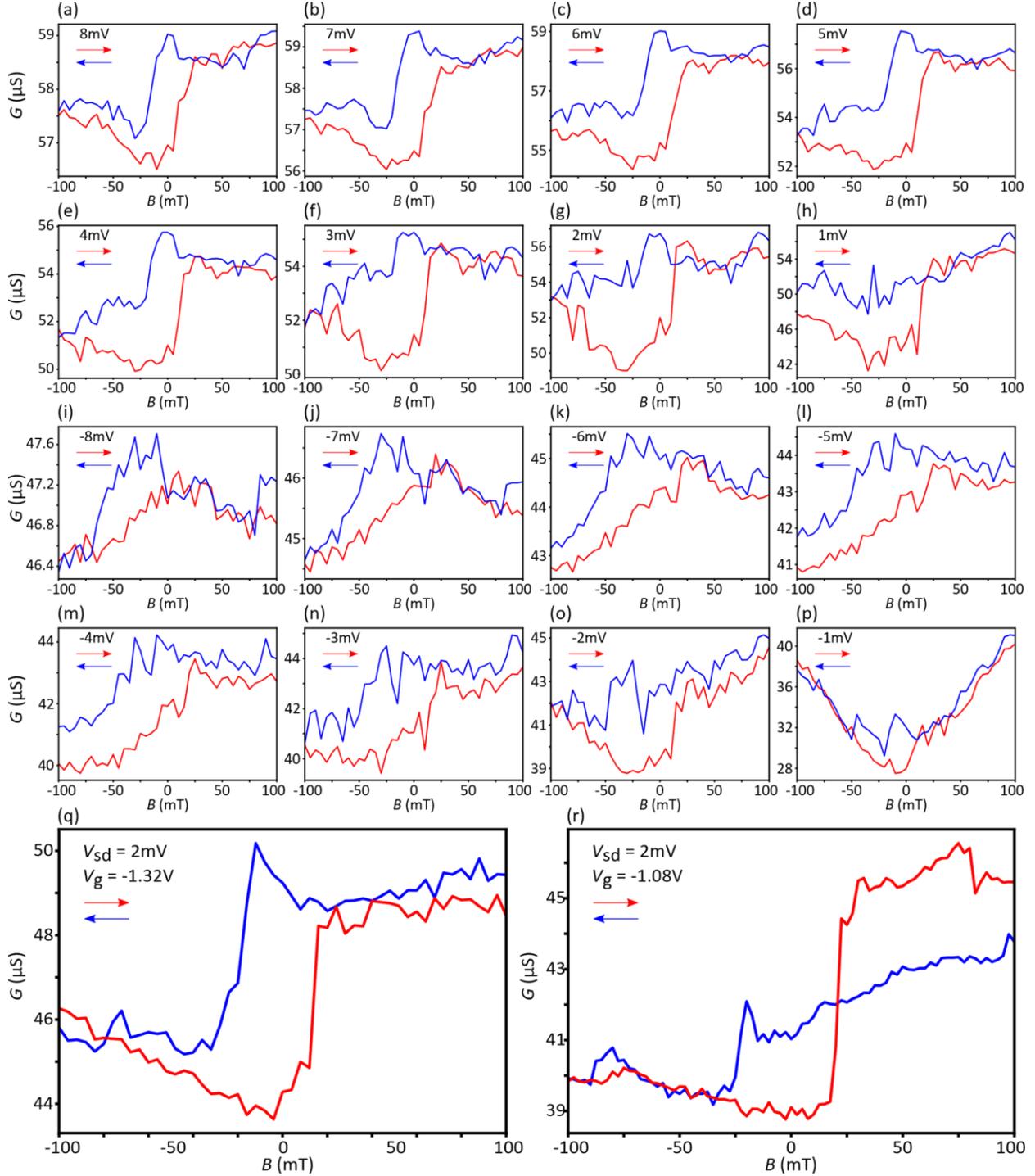

**Figure S8.** (a-h) Vertical line-cuts from Figs. 4 a,b from $V_{sd}$ = 8mV to 1mV. (i-p) Similar data for $V_{sd}$ = -8mV to -1mV. (q) Reproducibility of the filtering effect shown in a $G$ vs. $B$ measurement, with $B$ swept while $V_{sd}$ is kept constant at 2mV. (r) Reproducibility of the filtering effect after thermal cycling. The spin-filtering effect measurements were taken at $T = 100$mK.